\documentclass[12pt]{article} 
\usepackage{amssymb,amsfonts,amsmath,amsopn,amsthm} 
\usepackage[A4]{vmargin} 
\usepackage{calc} 
\usepackage{enumerate} 
 
\def\be {\begin{equation}} 
\def\ee {\end{equation}} 
 
\DeclareMathOperator*{\Div}{div} 
 
\DeclareMathOperator*{\const}{const} 
\DeclareMathOperator*{\identity}{identity}
\DeclareMathOperator{\mRe}{Re} 
\DeclareMathOperator{\mIm}{Im} 
 
\newcommand{\m}{\delta} 
\newcommand{\mb}{\bar{\delta}} 
\newcommand{\bm}{\mb}

\newcommand{\re}{\mathbb{R}} 
 
\newcommand{\Lie}{\pounds} 
\newcommand{\D}{\mathbf{D}} 
 
\newcommand{\IH}{\triangle} 
 
\newcommand{\kl}{\kappa^{(\ell)}}

\newcommand{\homega}{\hat{\omega}} 
 
\newcommand{\hnabla}{\hat{\nabla}} 
\newcommand{\hm}{\hat{m}} 
\newcommand{\hmb}{\overline{\hat{m}}} 
\newcommand{\hdiv}{\hat{\Div}} 
\newcommand{\hq}{\hat{q}} 
\newcommand{\hIH}{\hat{\triangle}} 
 
\newcommand{\e}{\epsilon} 
\newcommand{\tOmega}{\tilde{\Omega}} 
\newcommand{\laplaceS}{\m\mb+\mb\m-2a\m-2\bar{a}\mb} 
 
\newcommand{\tIH}{\tilde{\IH}}
\newcommand{\opD}{\ell^a \partial_a}      
 
\newtheorem{thm}{Theorem} 
\newtheorem{lem}{Lemma} 
\newtheorem{prp}{Proposition} 
\newtheorem{cor}{Corollary} 
 
\begin{document} 
 
\setcounter{secnumdepth}{2} 
 
\begin{center} 
\begin{minipage}[c]{\textwidth-3cm} 
\renewcommand*{\thempfootnote}{\arabic{mpfootnote}} 
\renewcommand*{\footnoterule}{} 
\title{Extremal Isolated Horizons: A Local Uniqueness Theorem} 
\author{Jerzy Lewandowski${}^{1,2,3}$, Tomasz Pawlowski${}^{1,2}$} 
\date{} 
\maketitle 
 
\footnotetext[1]{Instytut Fizyki Teoretycznej, Uniwersytet Warszawski,\\ 
   ul. Ho\.{z}a 69, 00-681 Warsaw, Poland} 
\footnotetext[2]{Max Planck Institut f\"ur Gravitationsphysik,\\ Albert  
  Einstein Institut, 14476 Golm, Germany} 
\footnotetext[3]{Physics Department, 104 Davey, Penn State, University 
  Park,\\ PA 16802, USA} 
\end{minipage} 
\end{center} 
\smallskip
\begin{center}\begin{large}
  Preprint: AEI-2002-082
\end{large}\end{center} 
\smallskip
 
\abstract{We derive all the axi-symmetric, vacuum, and  electrovac  
extremal isolated horizons. It turns out, that for every horizon in 
this class, the induced metric tensor, the rotation 1-form potential  
and the pullback of the electromagnetic field necessarily coincide with  
those induced by the monopolar, extremal Kerr-Newman solution on the 
event horizon. 
We also discuss the general case of a symmetric, extremal isolated horizon. 
In particular, we analyze the case of a two-dimensional symmetry group 
generated by two null vector fields. Its relevance
to the classification of all the symmetric isolated horizons, including the 
non-extremal ones, is explained.} 
 
\section{Introduction.}\label{sec:intro} 

\noindent{\bf Isolated horizons.} 
In a new model of black holes in equilibrium \cite{abf} the stationarity 
assumption is relaxed from spacetime and imposed only on an appropriately 
distinguished null surface. An isolated horizon (IH) $(\IH,\,[\ell])$ is 
a null surface $\IH$ equipped with a null flow $[\ell]$ which satisfies 
the following geometric and topological conditions ($\ell$ stands for a null 
vector field tangent to $\IH$ and generating the flow). The geometric 
conditions are:  
\begin{enumerate}[(i)] 
  \item the (degenerate) metric tensor $q$ induced on $\IH$ is preserved by 
    the flow $[\ell]$, and  
  \item the same holds for the covariant 
    derivative $\D$, induced on $\IH$\footnote{The first condition is 
      equivalent to a condition, that for every pair $X,Y$ of vector 
      fields tangent to $\IH$, the spacetime covariant derivative 
      $\nabla_XY$ is also tangent to $\IH$. The restriction of $\nabla$ 
      to the tangent bundle of $\IH$ is the induced covariant derivative 
      $\D$.} 
    \begin{subequations}\label{def}\begin{align} 
      \Lie_\ell q\ &=\ 0, & [\Lie_\ell, \D]\ &=\ 0. \tag{\ref{def}} 
    \end{align}\end{subequations} 
\end{enumerate} 
The topological condition is that $\IH$ contains 
a space-like 2-surface $\tIH$ homeomorphic to the 2-sphere, and that each 
orbit is homeomorphic to the line $\re$.    

In general there is no Killing vector field in a neighborhood of 
an isolated horizon \cite{l,abdfklw}, radiation may exist even arbitrarily
close to it. A local description of mechanics and geometry of IHs was 
developed in \cite{abf,abdfklw,abl-m,abl-g}. One of the structures  
used by this  description is the {\it IH geometry} defined by the pair 
$(q,\D)$. If one considers $\IH$ as a part of the characteristic 
Cauchy surface for the evolution of the gravitational field possibly coupled  
to some matter, then, in standard cases (Maxwell, scalar, 
Yang-Mills field, etc.) the pair $(q,\D)$ provides the part of the initial 
data corresponding to the  gravitational field. (The pair $(q,\D)$ might be 
thought of as the  spacetime metric tensor  given at $\IH$ up to the first 
order.) 
IH geometry has local degrees of freedom. The tensor $q$ is arbitrary
modulo the requirement that $\ell$ be orthogonal to $\IH$
and $\Lie_\ell q=0$. The ingredients  of $\D$ are listed in the next section.
Here we introduce one of them, namely a 1-form $\omega$ defined  
on $\IH$, such that   
\begin{equation} 
  \D\ell\ =\ \omega\otimes\ell, 
\end{equation} 
and called the {\it rotation 1-form potential}.
The existence of $\omega$ follows from $\Lie_\ell q=0$. The rotation 
of the IH is given by $d\omega$. The surface gravity $\kl$ of the vector  
field $\ell$  is  
\begin{equation} 
  \kl\ =\ \ell^a\omega_a. 
\end{equation} 
The surface gravity is  constant on $\IH$,  
\begin{equation}\label{0} 
  \kl\ =\ \const 
\end{equation} 
owing to the Zeroth Law \cite{abf}, which holds provided 
the Ricci tensor $R_{\mu\nu}$ satisfies at $\IH$  
\begin{equation} 
  \ell^a R_{a\mu}e^\mu_{(\IH)}\ =\ 0, 
\end{equation} 
where the subscript $(\IH)$ denotes the pullback of a covariant tensor  
onto $\IH$. In the Einstein-Maxwell case considered here, this 
condition follows from the constraint equations we discuss in the next  
section (in a more general context, this condition is implied by  
natural energy assumptions \cite{abl-g}).  
An IH $(\IH, [\ell])$ is called  {\it extremal}, whenever 
\begin{equation}\label{ekstr} 
  \kl\ =\ 0. 
\end{equation} 
This condition is independent of the choice of  generator 
of the flow $[\ell]$.  

The Einstein-Maxwell equations induce constraint equations on the horizon 
geometry  $(q,\D)$ and on the electromagnetic field possibly present on $\IH$.
The character of the equations crucially depends on the value of the 
surface gravity. Whereas the constraints reduce to a system of linear algebraic
equations in the {\it non}-extremal IH case \cite{abl-g}, in the 
extremal case  the constraints become a system of partial differential 
equations involving the 2-metric $q$, the rotation 1-form $\omega$ 
and an  electromagnetic field $F$  
(sections \ref{sec:constr:vac}, \ref{sec:constr:evac}). The general  
solution is not known. This is the issue we address in this paper.  

\medskip
\noindent{\bf Our results.}
The most surprising result contained in this paper follows
from our derivation of a general solution
of the vacuum and, respectively,  electrovac constraints  in the 
axi-symmetric, extremal IH case (section \ref{sec:axsol}).
We prove that if the electro-magnetic field  $F$ is also axi-symmetric 
on the horizon and preserved by the null flow $[\ell]$, then  $q$, $\omega$ 
and the pullback of $F$ onto $\IH$ coincide with those induced by the 
monopolar\footnote{The electromagnetic field depends on one more 
parameter, the possibly non-zero magnetic charge.}, 
extremal Kerr-Newman solution on the event horizon.
That uniqueness of the Kerr-Newman extremal horizon is surprising, because no 
non-local condition was imposed except that $\IH$ contains a slice 
homeomorphic to the 2-sphere. 
The question of what distinguishes the Kerr IH in a 
quasi-local theory was raised in \cite{lp1} and a local characterization
of the non-extremal Kerr IH was found. The current result 
completes that answer.    

Other results concern the general, possibly non-symmetric, electrovac, 
extremal IH case. We study integrability conditions for the constraints  
and their consequences 
(sections \ref{sec:constr:vac}, \ref{sec:constr:evac}). In particular, 
we demonstrate the non-vanishing of certain combinations of the Riemann 
tensor on $\IH$. The results are used to 
prove that, given a null 3-surface in spacetime, there is at most
one extremal IH structure\footnote{Of course, a generic null surface 
admits no IH structure.} defined on it, in the vacuum and 
electrovac cases  (section \ref{sec:uniq}). 
That issue was addressed in \cite{abl-g} and our current result completes 
the discussion of that paper.  

The above result is applied to the discussion of the symmetric extremal IHs 
(section \ref{sec:sym}). A map  $\IH\rightarrow \IH$ 
is called a symmetry, whenever it preserves the geometry $(q,\D)$, whereas 
the flow $[\ell]$ is not necessarily preserved. This definition 
is motivated by the properties of spacetime isometries:
every Killing vector field, if it exists, tangent to a given IH generates 
its symmetry.
An IH that admits a symmetry which is not an element of the flow $[\ell]$ 
is called symmetric.   
The cases we analyze are when the symmetry group contains
\begin{enumerate}[(i)]
  \item the O(2) group - the axi-symmetric case already mentioned in 
     this section and, respectively,  
  \item a two-dimensional group of null 
     symmetries - the {\it null-symmetric} case.   
\end{enumerate}

The null symmetric extremal IHs are also relevant for the classification 
of all symmetric IHs, including the non-extremal ones. Indeed, all the 
symmetric non-extremal IHs can be explicitly constructed by using the 
invariants introduced in \cite{abl-g} {\it  except} the case when the 
geometry $(q,\D)$ of a non-extremal $\IH$ admits an extra null symmetry. 
But then by \cite{abl-g}, one of the null generators of the symmetry group 
defines on the same $\IH$ another IH structure which is extremal, in the 
vacuum IH case. 
We generalize that result to the electrovac case in the current paper.
\medskip

\noindent{\bf Conventions.}
The spacetime signature is $(-,+,+,+)$. By the subscript $(\IH)$ we denote 
the pullback of a covariant tensor onto $\IH$. 
$(e^\mu)$, and $(e_\nu)$ stands for a coframe of 1-forms and 
a frame of vectors, respectively, null, unless indicated otherwise.
$e^1$ and $e^2$ are complex valued, $e^3$ and $e^4$ are real valued. 
The space-time metric is 
\begin{equation}
  g = e^1\otimes e^2 + e^2\otimes e^1 
    - e^3\otimes e^4 - e^4\otimes e^3.
\end{equation}
Due to the symmetry of every IH $\IH$, all the considerations
will automatically descend to the sphere $\hIH$ of the
null geodesics in $\IH$. We will not distinguish between
the functions defined on $\hIH$ and the functions defined
on $\IH$ which are constant along the null geodesics.

\section{The constraint equations}\label{sec:constr}  

In this section we discuss the 
constraint equations satisfied by the geometry $(q,\D)$ of an   
IH $(\IH, [\ell])$ in the vacuum and electrovac 
cases, respectively and their the integrability conditions. 
The constraints are given by that part of the 
spacetime energy-momentum tensor $T_{\mu\nu}$  
that can be derived from $(q,\D)$ via the Einstein equations, that is by 
the pullback of $T_{\mu\nu}$ onto $\IH$. The stress-energy   
tensor is constructed from an electro-magnetic  field present 
on $\IH$. We add to the system of equations, the 
constraints on the electromagnetic field $F$ on $\IH$ following from  
the Maxwell equations. The symmetry assumption is that the flow 
$[\ell]$ is a symmetry of the electromagnetic field on $\IH$.  
An arbitrary surface gravity is used in the equations to indicate 
the essential difference between the extremal and non-extremal case. 
The non-extremal case will be also considered in section \ref{sec:null:nex}.  
 
An IH $(\IH,[\ell])$ is called a {\it  vacuum IH}  
whenever the spacetime Ricci tensor $R_{\mu\nu}$ satisfies on $\IH$ 
\begin{equation}\label{G} 
  \left(R_{\mu\nu}e^\mu \otimes e^\nu\right)_{(\IH)}\ =\ 0. 
\end{equation} 
 
In the presence of an electromagnetic field $F$ at an IH 
$(\IH,[\ell])$, the part of the Einstein equations 
we assume on $\IH$  is the pullback 
\begin{equation}\label{GT} 
  \left((R_{\mu\nu}-8\pi T_{\mu\nu})e^\mu \otimes 
    e^\nu\right)_{(\IH)}\ =\ 0, 
\end{equation} 
where $T_{\mu\nu}$ is the gravitational energy-momentum tensor of 
the electromagnetic field, which satisfies in particular  
$T_\mu{}^\mu=0$. The Maxwell equations define extra constraints on the 
components of the electromagnetic field $F$ on $\IH$. We 
make a small effort now, to write them down in a geometric form. One of the 
constraints  on $F$ at $\IH$ is 
\begin{equation}\label{lFd} 
  \ell\lrcorner \star\left(d(F-i\star F)\right)\ =\ 0, 
\end{equation} 
where we used the spacetime Hodge star. For the differential forms defined 
on $\IH$  
and transversal to $\ell$, there is a well defined intrinsic Hodge dual  
${}^{(\IH)}\star $ (see the appendix). The remaining  constraint equation 
on $F$ is 
\begin{equation}\label{mF} 
  ({}^{(\IH)}\star\ -\ i)\left[\left(\star d(F-i\star F)\right)_{(\IH)}
\right]\ =\ 0. 
\end{equation}  
We will also assume that the electromagnetic field  
is invariant with respect to the null flow $[\ell]$. 
This assumption reads 
\begin{equation}\label{LielF} 
  \Lie_\ell \left(F-i\star F\right)_\nu{}^\mu e^\nu_{(\IH)}\otimes e_\mu \ =\ 
0, 
\end{equation} 
where the above Lie derivative is well defined for the following reasons. 
It follows from the equations \eqref{GT} and the vanishing 
of the expansion of $\ell$ that  
\begin{equation} 
  T_{ab}\ell^a\ell^b\ =\ 0. 
\end{equation} 
A consequence of this fact is 
\begin{equation}\label{lF} 
  \ell\lrcorner \left(F-i\star F\right)_{(\IH)}\ =\ 0. 
\end{equation} 
Therefore, if we consider the tensor $\left(F-i\star F\right)_\nu{}^\mu$  
as a 1-form  taking  vector values, then its pullback on $\IH$ takes values 
in the space $T(\IH)$ tangent to $\IH$.  
 
Concluding, an  
{\it electrovac isolated horizon} is an IH $(\IH, [\ell])$ 
which admits  an electromagnetic field $F$ such that 
the constraint equations \eqref{GT}-\eqref{mF}, \eqref{lF}  and 
the symmetry condition \eqref{LielF} are satisfied.
The electromagnetic field $F$ will be referred to as 
an electromagnetic field that {\it makes $\IH$ an electrovac IH}.  
 
Equation \eqref{lF} is the condition which implies 
that $\ell^a T_{a\mu}e^\mu_{(\IH)}=0$, the condition  
ensuring the Zeroth Law \eqref{0} via the constraint equations.  
   
\subsection{Geometry of an IH and the spacetime Ricci tensor}\label{sec:geom} 
\noindent{\bf Ingredients of $(q,\D)$ and $F_{(\IH)}$ in  
and adapted frame.}  
Before we turn to  the constraints on $(q,\D)$  in detail, 
we  specify the  elements  of $(q, \D)$ which are free
before imposing the constraints.
It follows from the definition that an isolated horizon $\IH$ can be viewed  
as a cylinder $\hIH \times \re$ generated by null geodesics where the 
projection 
\begin{equation}\label{Pi} 
  \Pi: \IH = \hIH \times \re \rightarrow \hIH 
\end{equation} 
maps each null geodesic into a point of the topological sphere $\hIH$. 
The induced metric tensor is 
\begin{equation} 
  q \ = \ g_{(\IH)} 
\end{equation} 
where $g$ is a space-time metric.

The assumption 
\begin{equation}  
  \Lie_\ell q \ = \ 0 
\end{equation} 
implies that $q$ is a pullback by $\Pi$ of a 2-metric tensor 
$\hq$ defined on $\hIH$, 
\begin{equation} 
  q\ =\ \Pi^*\hq. 
\end{equation} 
 
Let us fix a null vector field $\ell$ tangent to $\IH$ and representing 
the flow $[\ell]$. It is also convenient to  fix a foliation of $\IH$ by 
2-cross-sections (called 
`slices' hereafter) of the projection $\Pi$ preserved by 
the flow $[\ell]$. Then, a null vector 
field $n$ transversal to $\IH$ is uniquely defined by assuming the 
orthogonality to the slices and the normalizing condition 
\begin{equation}
  n_\mu \ell^\mu \ = \ -1. 
\end{equation} 
 
The covariant derivative $\D$ consists of the following 
components:  
\begin{enumerate}[(i)] 
  \item the metric connection defined on any slice by 
    the 2-metric $q$,  
  \item the rotation 1-form potential $\omega$, 
  \item the deformations of the induced metric tensor on a 
   2-slice while it is transported along the orthogonal null 
   vector field $n$. 
\end{enumerate} 
 
Let  $(e_1,\,e_2,\,e_3,\,e_4) = (m,\,\bar{m},\, n,\, \ell)$ be a null 
frame defined in a spacetime neighborhood of $\IH$. The degenerate metric 
tensor $q$ induced on $\IH$ is 
\begin{equation}
  q\ =\  (e^1 \otimes e^2 + e^2\otimes e^1)_{(\IH)}. 
\end{equation} 
The real vectors $\mRe(m)$, $\mIm(m)$ are (automatically) tangent 
to $\IH$.  To  adapt the frame further, we assume the 
vector fields $\mRe(m), \mIm(m)$ are tangent to the slices
of the fixed foliation. 
We can choose the vector field $m$ such that 
\begin{equation}\label{m} 
  \Lie_\ell m\ =\ 0, 
\end{equation} 
hence, the projection of $m$ onto $\hIH$ uniquely defines a null frame $\hm$ 
tangent to $\hIH$, 
\begin{equation} 
  \Pi_* m \ =:\ \hm. 
\end{equation} 
The frame $(e^1, e^2, e^3, e^4)$ is  adapted to: the vector field $\ell$,
the $[\ell]$ invariant foliation of $\IH$, and the null 
complex-valued frame $\hm$ defined on the sphere $\IH$.
Spacetime frames constructed in this way on $\IH$ will be called 
{\it adapted}.  

The elements of an adapted frame are Lie dragged by $\ell$, 
\begin{equation}
  \Lie_\ell e^\mu_{(\IH)}\ =  0. 
\end{equation} 
Therefore, the connection components defined by $\D$ in that frame 
are constant along the null geodesics tangent to $\ell$. 
They are, 
\begin{subequations}\begin{align} 
  \label{G12} 
    m^\nu\D \bar{m}_\nu\ &=\ 
    \Pi^* \left({\hm}^A\hnabla {\hmb}{}_A\right)\ =: \Pi^*\hat{\Gamma},\\ 
  \label{G43} 
    -n_\nu \D \ell^\nu\ &=\ \omega = 
    \pi e^2_{(\IH)} + \bar{\pi}e^1_{(\IH)} 
    + \kl e^3_{(\IH)},\\ 
  \label{G32} 
    -\bar{m}^\nu\D n_\nu\ &=\  \mu e^1_{(\IH)} +\lambda e^2_{(\IH)} 
    + \pi e^4_{(\IH)},\\ 
  m_\mu \D \ell^\mu\ &= \ 0, 
\end{align}\end{subequations} 
where $\hat{\Gamma}$  is the Levi-Civita connection 
1-form corresponding to the covariant derivative $\hat{\nabla}$
defined by $\hq$ and to the null frame  $\hm$ defined on $\hIH$.

Due to $[\Lie_\ell, \D]\ =\ 0$,  
\begin{equation} 
  \Lie_\ell \omega\ =\ 0. 
\end{equation} 
Therefore, there is  a 1-form $\hat{\omega}$ defined on 
$\IH$, such that 
\begin{equation}\label{homega} 
  \Pi^*\hat{\omega}\ =\  \pi e^2_{(\IH)} + \bar{\pi}e^1_{(\IH)}. 
\end{equation} 
In the case  of an extremal IH, $\homega$ is uniquely defined on $\IH$
by $[\ell]$ and $\D$. 

The  global  existence  of solutions to the constraint equations  
on $\IH$ will be crucial for us, hence we conclude this paragraph  
with discussion of the global existence of the objects defined above.   
Defined globally 
on $\IH$ are: the vector field $\ell=e_4$, the covector $e^4_{(\IH)}$,  
the function $\mu$, the tensor $\lambda e^2\otimes e^2$, the 1-forms  
$\omega$ and $\homega$, and the constant $\kl$. 
It will be convenient to express the 1-form $\homega$ 
by two functions  $U$ and $\ln B$ globally  defined on $\hIH$ 
(sufficiently many times differentiable) using the well known 
decomposition 
\begin{equation}\label{UBa} 
  \homega\ =\ \hat{\star}dU + d\ln B
\end{equation} 
where $\hat{\star}$ is Hodge star defined by the 2-metric
tensor $\hq$.
The coefficient $\pi$ of $\homega$ becomes 
\begin{equation}\label{pi} 
  \pi \ =\ -i\bar{\delta}U + \bar{\delta}\ln B. 
\end{equation}  
\medskip 
 
\noindent{\bf The Ricci tensor components present in the constraints.}  
We turn now to the pullback of the spacetime Ricci tensor  onto  
$\IH$ and express it by $(q,\D)$. We still admit a possibly 
non-zero  surface gravity 
$\kl$ both because the non-extremal case will be considered in the last  
section, and to emphasize the difference between the extremal 
and non-extremal cases. 
 
Due to $\Lie_\ell q=0$, $\Lie_\ell\omega=0$, 
$d\kl=0$, and $[\Lie_\ell,\D]=0$ the Ricci tensor already satisfies   
\begin{subequations}\label{eq:ricci}\begin{align}
  \ell^a R_{a\mu}e^\mu_{(\IH)}\  &=\ 0, &   
  \Lie_\ell \left(R_{\mu\nu}e^\mu\otimes e^\nu\right)_{(\IH)}\ &=\ 0. 
    \tag{\ref{eq:ricci}} 
\end{align}\end{subequations} 
The remaining components of the pullback onto $\IH$ of the 
Ricci tensor will be expressed in terms of the following objects  
defined on the sphere $\hIH$: the null frame $(\hat{m}, \bar{\hat{m}})$  
and the dual 
coframe $(\hat{e}^1,\hat{e}^2 )$, the complex valued differential 
operator 
\begin{equation}
  \delta := \hm^A\partial_A, 
\end{equation} 
(where $A=1,2$ refers to coordinate system $(x^1, x^2)$ defined 
locally on $\hIH$), the connection form $\hat{\Gamma}$ defined above 
\begin{equation}\label{2gamma} 
  \hat{\Gamma}\ =:\ 2\bar{a} \hat{e}^1 + 2{a} \hat{e}^2; 
\end{equation} 
its curvature scalar
\begin{equation}\label{k} 
  K := 2\m a  + 2\bar{\m} \bar{a} - 8 a  \bar{a}; 
\end{equation} 
and the 1-form $\hat{\omega}$ and its divergence 
\begin{equation}\label{div} 
  \hdiv \hat{\omega}\ = \m\pi + \bm\bar{\pi} - 
  2a\bar{\pi} - 2\bar{a}\pi. 
\end{equation} 
 
The spacetime Ricci tensor components  in question are  
\begin{subequations}\label{D}\begin{align} 
  \label{Dmu} 
    R_{m\bar{m}}\ &=\ 2\kl \mu - \hdiv \homega - 
    \homega^2 + K ,\\ 
  \label{Dlambda} 
    R_{\bar{m}\bar{m}}\  &=\  2\kl \lambda - 2\mb \pi - 4a\pi - 
    2\pi^2, 
\end{align}\end{subequations} 
where $\homega^2 := \homega_A \homega_B \hq^{AB}$. 
As we mentioned before, the components have to be constant 
along the null geodesics.  Note that  
the remaining connection components $\mu$ and $\lambda$ 
are not involved in the formula for the pullback of the 
Ricci tensor. 
\medskip 
  
\subsection{The constraints in the extremal vacuum case}\label{sec:constr:vac} 
We consider now an extremal, vacuum IH; that is, we
assume equations \eqref{ekstr}, \eqref{G} hold on $\IH$. 
The constraints amount to the following equations on the  metric $\hq$ and the 
1-form  $\homega$ defined on the sphere $\hIH$:
\begin{subequations}\begin{align} 
  \label{1} 
    \hdiv \homega + \homega^2 - K\ &=\ 0,\\ 
  \label{2} 
    \mb \pi + 2a\pi + \pi^2 \ &=\ 0. 
\end{align}\end{subequations} 
This system of equations can also be written in the following
compact, tensorial form
\begin{equation}
\hat{\nabla}_{(A}\homega_{B)}\ +\ \homega_A\homega_B\  - 
\frac{1}{2}\hat{R}_{AB}\ =\ 0, 
\end{equation} 
where $\hat{\nabla}$ and $\hat{R}_{AB}$ are the covariant derivative
and Ricci tensor defined on the sphere $\hIH$ by the 2-metric $\hq$. 
 
In terms of the functions $U$ and $B$ defined in \eqref{UBa}, \eqref{pi}, 
equations \eqref{1}, \eqref{2} read as  
\begin{subequations}\label{vac}\begin{align} 
  \big(\m\mb + \mb\m - 2a\m - 2\bar{a}\mb + 2i \m U \mb - 2i \mb U \m 
     + 2 \m U \mb U - K \big) B \ &=\ 0,\\ 
  \big(\mb\mb + 2a\mb -2i \mb U\mb -i\mb\mb U -2ia\mb U - \mb U\mb U  \big)B 
     \ &=\ 0, 
\end{align}\end{subequations} 
where by $\m U$ or $\m \m U$, etc we mean  functions, not the 
product of the operators. Note that the equations do not
involve the real-valued  scalar $\mu$ and 
the complex-valued tensor $\lambda\hat{e}^2\otimes \hat{e}^2$, which are  
therefore freely defined on $\hIH$. An equivalent system of equations was 
considered  in \cite{abl-g}.  
An integrability condition given by commuting the $\delta$ 
operator with  $\m\mb + \mb\m$ as well $\mb$ with $\m\m$  
is 
\begin{equation} \label{bian} 
  \mb\Psi_2 + 3\pi\Psi_2\ =\ 0  
\end{equation} 
where  $\Psi_2$ is an invariant which characterises the geometry $(q,\D)$ 
of a vacuum IH, 
\begin{equation} 
  \Psi_2\ =\ \frac{1}{2}\left( -K + i\left( \laplaceS \right)U \right).  
\end{equation} 
From the spacetime point of view,
the function $\Psi_2$ is the spinorial  component $C_{\ell n \ell n}- 
C_{\ell n m \bar{m}}$ of  
the Weyl tensor  
$C_{\mu\nu\alpha\beta}$, it is always constant along the null generators of 
$\IH$
and the integrability condition \eqref{bian} 
is one of the Bianchi identities. The condition implies the  
following identity via  \eqref{pi}, 
\begin{equation}\label{psi2b}  
  \Psi_2 B^3 e^{-3iU}\ = \ C_0=\const.  
\end{equation} 
Since  $B$ nowhere vanishes (by definition \eqref{UBa}), and $K\neq 0$ 
on a nontrivial subset of $\hIH$, we have: 
\begin{equation}\label{eq:vac_C} 
  C_0 \neq 0.  
\end{equation} 
The solution for $B$ is 
\begin{equation} 
  B \ =\ B_0 \left(\Psi_2\right)^{-\frac{1}{3}}e^{iU} 
\end{equation}  
where  $B_0\not= 0$ is a constant. 
The conditions 
\begin{equation} 
  B\ =\ \bar{B} \ >\ 0 
\end{equation}  
imply  constraints on the remaining unknowns, that is on the metric tensor  
$q$ and the function $U$. Another  consequence of (\ref{eq:vac_C}) that will 
be important in section \ref{sec:uniq} is the following 
{\it non-vanishing lemma}:  
\medskip

\begin{lem}\label{thm:vac} 
 For every extremal, vacuum IH
the invariant $\Psi_2$ nowhere vanishes on $\IH$:
\begin{equation}\label{nonvan} 
  \Psi_2(x)\ \neq\ 0 
\end{equation}  
for every $x\in \IH$. 
 \end{lem} 
\medskip 
 
\subsection{The constraints in the extremal electrovac case}
\label{sec:constr:evac} 
Now we consider an extremal, electrovac IH; that is, we  assume  
that equations \eqref{ekstr}, \eqref{GT}-\eqref{LielF}
hold on $\IH$. 
Let $F=\frac{1}{2}F_{\mu\nu}e^\mu\wedge \e^\nu$ be an 
electromagnetic field present in a spacetime neighborhood of the 
horizon. We use the standard notation for the components of $F$ in 
a null frame, 
\begin{equation}\label{F} 
  F\ =\ -\Phi_0e^4\wedge e^1 + \Phi_1(e^4\wedge e^3 + e^2\wedge e^1) 
  - \Phi_2 e^3\wedge e^2 + c.c.. 
\end{equation} 
Condition \eqref{lF} implied by the constraints reads 
\begin{equation}\label{Phi0} 
  \Phi_0 \ =\ 0. 
\end{equation} 
 
The assumed invariance of $F_\nu{}^\mu e^\nu_{(\IH)}\otimes e_\mu$ 
with respect to the null flow $[\ell]$ is equivalent to the 
requirement that the coefficients $\Phi_1$ and $\Phi_2$ be 
constant along the null geodesics generating $\IH$. 
 
Now, the part of the constraints \eqref{lFd}, \eqref{mF} coming from the 
Maxwell equations amounts to the following equation on $\Phi_1$ only, 
\begin{equation}\label{mbphi} 
  \mb \Phi_1  + 2\pi \Phi_1\ =\ 0, 
\end{equation} 
which is easy to integrate into the following form: 
\begin{equation}\label{phi} 
  \Phi_1\ =\ E_0 B^{-2} e^{2i U}, 
\end{equation} 
$E_0$ being a complex constant, and the functions $U$ and $B$ 
being  defined  by \eqref{pi}.  
 
The part of the constraints following from the Einstein equations 
\eqref{GT} becomes
\begin{subequations}\label{eq:_constr1}\begin{align}
  R_{mm}\ &=\ 0, & 
  R_{m\bar{m}}\ &=\ 4|\Phi_1|^2. \tag{\ref{eq:_constr1}} 
\end{align}\end{subequations} 
In terms of the geometry $\hq$ of the sphere $\hIH$
and the 1-form $\homega$, the constraints read 
\begin{equation}
\hat{\nabla}_{(A}\homega_{B)}\ +\ \homega_A\homega_B\  - 
\frac{1}{2}\hat{R}_{AB}\ +\ 2|\Phi_1|^2\hq_{AB}\ =\ 0. 
\end{equation} 
Finally, after expressing $\homega$ by the the functions $U$ and $B$,
the constraints read, 
 
\begin{subequations}\label{112}\begin{align} 
  \label{12} 
    \begin{split} 
      \left(\m\mb + \mb\m - 2a\m - 2\bar{a}\mb + 2i \m U \mb - 2i \mb U \m 
            + 2 \m U \mb U - K  \right) B\ + \\ 
      + 4\frac{|E_0|^2}{B^3} \ &=\ 0, 
    \end{split}\\ 
  \label{11} 
    \big(\mb\mb + 2a\mb -2i \mb U\mb -i\mb\mb U -2ia\mb U - \mb U\mb U  \big)B 
    \ &=\ 0. 
\end{align}\end{subequations} 
Concluding, an extremal IH $(\IH, [\ell])$ 
is  an extremal, electrovac IH if and only if it admits
an electromagnetic field  $F$ (\ref{F}) such that the components 
$\Phi_1, \Phi_2$ are constant along the null geodesics in $\IH$ and
the geometry $(q,\D)$ and $F$ satisfy equations 
\eqref{Phi0}, \eqref{phi} and \eqref{112}. Note that the components
$\mu, \lambda$ of $\D$ as well as the component $\Phi_2$ of $F$
are not involved in the equations, and $\Phi_1$
is a complex-valued function defined globally on the sphere $\hIH$. 
The assumption that $\Phi_2$ be constant along the null geodesics 
in $\IH$ was used to eliminate the derivatives $\opD\Phi_2$ from 
the Maxwell equations and derive the equation \eqref{mbphi}. 
 
We complete this paragraph by a discussion of consequences 
of the constraint equations. 
It follows from \eqref{phi}, that:

\begin{lem}\label{el} 
The component $\Phi_1$ of the electromagnetic field either vanishes nowhere 
on $\IH$, or it is zero everywhere on $\IH$. In the second case, 
$\IH$ is a vacuum isolated horizon (even if $\Phi_2\not=0$).  
\end{lem} 

Commuting  the $\delta$ operator with  $\m\mb + \mb\m$ as well $\mb$  
with $\m\m$ we derive the following integrability conditions 
for the equations \eqref{112} 
\begin{equation}\label{eq:Psi2_em} 
  (\mb + 3\pi)\Psi_2 - \pi R_{m\bar{m}}\ =\ 0, 
\end{equation} 
where  as before, $\Psi_2$ is the component $C_{4343}-C_{4312}$ 
of the Weyl tensor, and in the presence of the electromagnetic field  
it takes on $\IH$ the following value: 
\begin{equation}\label{b} 
  \Psi_2 \ = \ 2\frac{|E_0|^2}{B^4} 
             + \frac{1}{2}\left( - K + i\left(\laplaceS\right)U \right).  
\end{equation}   
The integrability condition is one of the Bianchi identities. 
 
In this case we were not able to solve completely the integrability 
condition for the functions $\Psi_2$ and $B$. However, the fact of 
non vanishing of $\Psi_2$ on an extremal vacuum IH can be generalized 
in the following way (applied later in section \ref{sec:uniq}):  
\medskip 
 
\begin{lem}\label{thm:lemma} 
  For every extremal electrovac IH $\IH$, there is an open 
  and dense subset  ${\cal V}\subset \IH$ such that 
  \begin{equation} 
    \left. 3\Psi_2 - {R_{m\bar{m}}} \right|_{\cal V} \not=\ 0. 
  \end{equation} 
\end{lem} 
 
\medskip  
 
\begin{proof}[Proof] 
In the vacuum case the conclusion of lemma was already proved, 
hence we prove it now assuming that $R_{m\bar{m}}$ is not identically  
zero on $\IH$. As a consequence, $R_{m\bar{m}}$ is nowhere zero;  
that is, for every point $x\in \IH$, 
\begin{equation}\label{ass} 
  R_{m\bar{m}}(x)\ \not=\ 0. 
\end{equation} 
 
Using equations 
\eqref{phi} and \eqref{eq:Psi2_em} we derive the following 
equation   
\begin{equation}\label{eq:Psi2_B} 
  \mb\left( (3\Psi_2-R_{m\bar{m}}) B^3 e^{-3iU} \right) 
    + 16|E_0|^2 e^{-3iU} \mb B^{-1} = 0. 
\end{equation} 
where  
\begin{equation}  
  E_0\not=0, 
\end{equation} 
due to the assumption (\ref{ass}). 
Let ${\cal U}$ be an open subset of $\IH$ on which 
\begin{equation}\label{7} 
  3\Psi_2 - R_{m\bar{m}}\ =\ 0. 
\end{equation}  
It follows from \eqref{eq:Psi2_B} that  
\begin{equation} 
  B |_{\cal U} = \const \not= 0. 
\end{equation} 
Then, equation \eqref{12} takes the following form 
\begin{equation} 
  \left. -K + R_{m\bar{m}} + 2|\mb U|^2 \right|_{\cal U}\ = \ 0. 
\end{equation} 
Moreover, it follows from \eqref{7} that  
\begin{equation} 
  K |_{\cal U}\ =\ \frac{1}{3} R_{m\bar{m}} 
\end{equation} 
hence, we have finally 
\begin{equation} 
  \left. R_{m\bar{m}} + 3|\mb U|^2 \right|_{\cal U} = 0. 
\end{equation}  
Since each of the terms is non-negative, they both vanish  
on ${\cal U}$.  This contradiction completes the proof.  
\end{proof} 
\medskip 
 
\subsection{Uniqueness of the electrovac extremal IH}\label{sec:uniq} 
Given an extremal  IH $(\IH, [\ell])$ and its geometry $(q,\D)$ a natural  
question is whether there exists another null flow  
\begin{equation} 
  [\ell']\ \neq\ [\ell] 
\end{equation} 
such that $(\IH, [\ell'])$ is an extremal IH as well.  
Suppose then that there exists a function $f$ defined on $\IH$ such that  
\begin{subequations}\label{extr}\begin{align} 
  \ell' &= f\ell,          &   
  [\Lie_{\ell'},\D] &= 0,  &   
  \D_{\ell'}\ell'&=0.      \tag{\ref{extr}} 
\end{align}\end{subequations} 
The first two equations are equivalent to the following 
equation \cite{abl-g}: 
\begin{equation}\label{nulsym} 
  \left(\D_a\D_b + 2\omega_{(a}\D_{b)}\right)f\ =\ 0. 
\end{equation} 
The last equation in \eqref{extr} implies that the function $f$ is constant 
along the null geodesics tangent to $\ell$,  
\begin{equation} 
  \opD f \ =\ 0. 
\end{equation} 
The integrability condition for \eqref{nulsym} obtained by acting 
on it with $\D_a$ can be expressed by the spacetime curvature 
components \cite{abl-g} 
\begin{equation} \label{int1} 
  \left(3\Psi_2 - R_{m\bar{m}}\right)\mb f\ = 0,  
\end{equation} 
where we have already assumed that $\IH$ is an electrovac IH. 
Owing to lemma \ref{thm:lemma}, the condition above implies   
$f=\const$, and  
\begin{equation} 
  [\ell'] = \ell. 
\end{equation} 
 
\medskip 
 
\begin{prp}\label{thm:ExUniq} 
  If $(\IH, [\ell])$ is an extremal electrovac 
  IH (including the vacuum case) and a null vector field $\ell'$ 
  tangent to $\IH$ satisfies  
  \begin{subequations}\label{eq:prp1}\begin{align} 
    \Lie_{\ell'} q\ &=\ 0,   & 
    [\Lie_{\ell'}, \D]\ &=\ 0,   &
    D_{\ell'}\ell'\ &=\ 0;   \tag{\ref{eq:prp1}} 
  \end{align}\end{subequations} 
  Then $[\ell'] = [\ell]$. 
\end{prp}  
\medskip 
 
In conclusion, given a null surface $\IH$ in a spacetime, there is at  
most one extremal IH $[\ell]$ defined thereon.     
\medskip 
 
\section{Symmetric IHs}\label{sec:sym} 
Given an IH $(\IH,\ell)$ suppose that there is a vector field 
$X$ tangent to $\IH$,  such that   
\begin{equation}\label{axi} 
  \Lie_X q\ =\ 0\ =\ [\Lie_X, \D]. 
\end{equation} 
We call $X$ a generator of a  symmetry of the geometry $(q,\D)$. 
Obviously, the vector field $\ell$ is a symmetry itself. 
If $X \notin [\ell]$, then we call  $\IH$  a symmetric IH. 
Note that it is not {\it assumed} that the symmetries 
preserve $[\ell]$. Nonetheless they do so, however, at least 
in the extremal, electrovac or vacuum case, due to proposition 
\ref{thm:ExUniq}.  
\medskip 
 
\begin{cor}\label{thm:cor} 
  Given an electrovac extremal IH $(\IH,[\ell])$,  
  suppose $X$ is a  generator of a symmetry of the geometry $(q,\D)$.  
  Then, the flow $[X]$ preserves the flow $[\ell]$, that is 
  \begin{subequations}\label{eq:_col1}\begin{align} 
    [X,\ell]\ &=\ a_0\ell,  &  a_0\ &=\ \const. \tag{\ref{eq:_col1}}   
  \end{align}\end{subequations} 
\end{cor}  
In this section we discuss two cases of symmetric extremal isolated 
horizons: the axisymmetric case and the null-symmetric case.
Since the geometry $(q,\D)$ of a null-symmetric extremal electrovac IH 
admits  also a non-extremal IH structure, for  completeness we
formulate in the last subsection a converse statement. 

\subsection{Axisymmetric extremal IHs.} 
Consider an extremal IH $(\IH,[\ell])$ whose geometry $(q,\D)$   
has a symmetry  group isomorphic to O(2) and which satisfies
the conclusion of corollary \ref{thm:cor}.
We show now that in this  
case, the symmetry group  $O(2)$ commutes with the flow 
$[\ell]$, and we adapt coordinates on $\IH$. The constraint equations 
will be solved in the next section.

 Label the elements of the 
group by a parameter,   $[0,2\pi]\ni \varphi \mapsto\ U(\varphi)$ 
such that  
\begin{subequations}\label{eq:_ax}\begin{align} 
  U(\varphi_1)U(\varphi_2)\ &=\ U(\varphi_1+\varphi_2), &  
  U(0)=U(2\pi) &= \identity.  \tag{\ref{eq:_ax}}  
\end{align}\end{subequations} 
According to corollary \ref{thm:cor}, a vector field $U(\varphi)_*\ell$  
defines the same IH as $\ell$, hence 
\begin{equation} 
  U(\varphi)_*\ell\ =\ a_0(\varphi)\ell, 
\end{equation}   
where $a_0(\varphi)\in\re$ is a constant. As a function of $\varphi$, it 
satisfies 
\begin{subequations}\label{eq:_ax2}\begin{align} 
  a_0(\varphi_1)a_0(\varphi_2)\ &=\ a_0(\varphi_1+\varphi_2), &  
  a_0(0) &= a_0(2\pi) = 1.  \tag{\ref{eq:_ax2}}  
\end{align}\end{subequations} 
The only solution is  
\begin{equation} 
  U(\varphi)_*\ell\ =\ \ell. 
\end{equation}   
Therefore, the following is true: 
  
\begin{prp}\label{thm:ExAx} 
  Suppose $(\IH, [\ell])$ is an extremal, 
  electrovac IH; suppose further a vector field ${\bf \Phi}$ tangent to $\IH$  
  is a generator of a proper symmetry group of the IH geometry, 
  and the group  it generates is isomorphic to O(2). Then    
  \begin{equation} 
    \Lie_\ell{\bf \Phi}\ =\ 0. 
  \end{equation} 
\end{prp} 
\medskip 
 
Given an axisymmetric IH, we adjust coordinates on $\IH$ 
in the following way. Notice first, that due to proposition \ref{thm:ExAx} 
 the projection $\Pi_*{\bf \Phi}$ defines 
uniquely a vector field on $\hIH$, 
\begin{equation}
  \hat{\bf \Phi}\ := \Pi_*{\bf \Phi} 
\end{equation} 
and the first equality in \eqref{axi} implies that 
\begin{equation}
  \Lie_{\hat{\Phi}} \hq\ =\ 0. 
\end{equation} 
Therefore $\hat {\bf \Phi}$ indeed generates a group of rotations 
of $\hq$. It follows that there are spherical coordinates $(\theta, \varphi)$ 
on $\hIH$ naturally pulled back to $\IH$,  and an extra coordinate $v$ on  
$\IH$, such that 
\begin{equation}
  {\bf \Phi}\ =\ a(\theta, \varphi)\partial_v + \partial_\varphi, 
\end{equation} 
modulo a constant factor. To kill the first term locally, 
it is enough to replace $v$ by a new coordinate 
\begin{subequations}\label{eq:_ax3}\begin{align}
  v' \ &=\ v - A(\theta,\varphi),  & 
  A(\theta,\varphi)\ &:=\ \int^\varphi_{\varphi_0(\theta)} a(\theta,\varphi'). 
    \tag{\ref{eq:_ax3}} 
\end{align}\end{subequations} 
The resulting $v'$ is a continuous function defined globally 
on $\IH$, due to the condition 
\begin{equation}
  \int_{\varphi_0}^{\varphi_0+2\pi} a(\theta,\varphi)\ =\ 0, 
\end{equation} 
which follows from the assumption that the orbits of $\Phi$ are 
closed and from the fact that each orbit can intersect each null  
geodesics in $\IH$ only once (otherwise, the orbit would have a 
self-intersection). The differentiability of $v'$ may be ensured by 
appropriate choice of the boundary of integration 
$\varphi_0(\theta)$ as a function of $\theta$. In the coordinates 
$(v',\theta,\varphi)$, the vector field ${\bf \Phi}$ is 
\begin{equation}
  {\bf \Phi}\ =\ \partial_\varphi. 
\end{equation} 
Instead of the auxiliary coordinate $\theta$ on $\hIH$, used 
implicitly above, we introduce a `Hamiltonian' function, naturally defined by 
the vector field $\hat{\Phi}$ and the area 2-form $\hat{\epsilon}$ of $\hq$, 
namely a function $x$ such that 
\begin{equation}
  \hat{\Phi} \lrcorner \hat{\epsilon} \ =\ 2dx. 
\end{equation} 
The global existence on $\hIH$ and the differentiability of this function 
is ensured by the fact that 
\begin{equation}
  d \big( \hat{\Phi} \lrcorner \hat{\epsilon} \big)\ =\ 
  \Lie_{\hat{\Phi}} \hat{\epsilon}\ =\ 0. 
\end{equation} 
The function $x$ has exactly two extremal points, a minimum and 
a maximum, at the two zero points of $\hat{\Phi}$, and the additive 
constant is fixed by the condition 
\begin{subequations}\label{eq:_ax4}\begin{align}
  \sup x \ &=\ \frac{A}{8\pi}, & 
  \inf x \ &=\ - \frac{A}{8\pi}, \tag{\ref{eq:_ax4}} 
\end{align}\end{subequations} 
where $A$ is a constant, equal to the area of $\hIH$. 
Concluding, we are going to use the functions $(\varphi,x)$ 
as spherical coordinates on $\hIH$, and, denoting in the same way  their 
pullbacks to $\IH$, the functions $(v', \varphi, x)$ as coordinates on $\IH$. 
Adjusting the 2-frame  $\hat{m}$ on $\hIH$ such that 
\begin{equation}
  \Lie_{\hat{\Phi}} \hat{m}\ =\ 0, 
\end{equation} 
we get the adapted frame $(e^1, e^2, e^3, e^4)$ whose pullbacks 
onto $\IH$ are invariant with respect to the rotations generated by 
$\Phi$. 
\medskip 
  
\subsection{Null-symmetric extremal IHs}\label{sec:null:ex} 
An IH is called null-symmetric if it admits a  generator 
$X$ of a symmetry, such that  
\begin{subequations}\label{eq:_null1}\begin{align} 
  q_{ab}X^aX^b \ &= \ 0,  & 
  X \ &\notin \ [\ell].  \tag{\ref{eq:_null1}} 
\end{align}\end{subequations} 
Suppose this is the case. There exists a function $f$ defined 
on $\IH$, such that  
\begin{equation} 
  X \ =\ f\ell , 
\end{equation}  
and the equations \eqref{nulsym} hold. It follows from them that 
\begin{equation}\label{Df} 
  \D_b \opD f\ =\ 0.  
\end{equation}  
If $v$ is a function defined on $\IH$, such that 
\begin{equation} 
  \opD v\ =\ 1, 
\end{equation} 
then the general solution of \eqref{Df} is 
\begin{subequations}\label{f}\begin{align} 
  f\ &=\ b + \kappa_0 v, & 
  \opD b \ &= \ 0, & 
  \kappa_0 \ &= \ \const \neq 0. \tag{\ref{f}} 
\end{align}\end{subequations} 
The case of $\kappa_0=0$ is excluded by proposition 
\ref{thm:ExUniq}. Note that every linear combination $X-a_0\ell$, 
$a_0$  a constant, defines a non-extremal IH $(\IH, [X])$ 
(that is, the null surface $\IH$ admits many distinct IHs). 
Therefore \cite{abl-g}, every cross section of $\IH\rightarrow \hIH$ 
given by the equation 
\begin{equation}\label{fol} 
  b + \kappa_0 v \ = \ \const, 
\end{equation}  
has identically vanishing shear and expansion in each 
null direction orthogonal to this cross section. In other words, 
if we use a null frame $(m,\bar{m},n,\ell)$ adapted to the  
foliation of $\IH$ given by \eqref{fol}, then the  
following coefficients of $\D$ vanish (see \eqref{G32})    
\begin{equation} 
  \mu \ = \ \lambda \ = \ 0. 
\end{equation} 
 
The converse is also true. Suppose an extremal IH $(\IH, [\ell])$ 
admits  a foliation by 2-space-like surfaces such that 
every null vector field tangent to any leaf has zero shear and expansion 
on $\IH$. Then \cite{abl-g} $\IH$ is a null-symmetric IH. The proper null 
symmetry 
is generated by the vector field  
\begin{equation} 
  X\ =\ v\ell  
\end{equation}    
where $v$ is a function constant on the leaves and such that  
$\ell^av_{,a}=1$. 
 
Even though there are still some degrees of freedom left in the
geometry of a null-symmetric extremal IH $(\IH,[\ell])$, 
the group of the null symmetries is always the same in the electrovac case.
We characterize this group now.  

It follows from \eqref{f} that the Lie algebra defined by the vector  
fields $\ell$ and $X$ above is unique, namely given by the
following commutator
\begin{equation}\label{alg} 
  [\ell,\frac{1}{\kappa_0}X]\ =\ \ell.  
\end{equation} 
Suppose $X'$ is another null generator of a null symmetry.
We will see it belongs to the Lie algebra spanned by $\ell$ and $X$. 
Indeed, as before 
\begin{subequations}\label{eq:three}\begin{align} 
  X'\ &=\ (b' +\kappa'_0v)\ell, & 
  \opD b'\ &=\ 0, & 
  \kappa'_0\ &\neq\ 0. \tag{\ref{eq:three}} 
\end{align}\end{subequations} 
Consider the following linear combination of the generators, 
\begin{equation} 
  Y\ :=\ \frac{1}{\kappa_0}X-\frac{1}{\kappa'_0}X'.  
\end{equation} 
It is also a null generator of a symmetry of $(q,\D)$. Note that 
\begin{equation} 
  Y\ =\ \left(\frac{1}{\kappa_0}b-\frac{1}{\kappa'_0}b'\right)\ell 
  \ =\ b''\ell,\ \ \opD b''\ =\ 0.  
\end{equation} 
Therefore, 
\begin{equation} 
  \D_Y Y\ =\ 0. 
\end{equation} 
Owing to proposition \ref{thm:ExUniq}, if $(\IH,[\ell])$ is an extremal  
electrovac IH, then there is a constant $a'_0$ such that 
\begin{equation} 
  Y\ =\ a'_0\ell. 
\end{equation} 

In conclusion: 

\begin{prp}
A general null-symmetric electrovac (including vacuum) extremal IH $(\IH, 
[\ell])$ is given by any  solution $(\hq, \homega)$ of the 
constraint equations \eqref{12}, \eqref{11} and $\mu=0=\lambda$ in 
\eqref{G32}. 
Its group of the null symmetries is exactly two-dimensional, the generators 
are a vector field
$\ell\in [\ell]$ and $X=v\ell$ where $v$ is a function constant
on the foliation corresponding to $\mu$ and $\lambda$,
and such that $\ell^a v_{,a} = 1$. The commutator between
the generators is
\begin{equation}
[\ell,X]\ =\ \ell.
\end{equation}  
\end{prp} 

\subsection{Null-symmetric non-extremal IHs}\label{sec:null:nex} 
As noted above, a null-symmetric, extremal, electrovac IH
admits a null vector field which defines on the same null
surface a non-extremal IH. In this subsection, we 
discuss a converse statement.
  
Consider a non-extremal IH  $(\IH, [\ell])$ 
such that there exists on $\IH$ a null vector field 
$X$ whose flow is different from $[\ell]$ and which generates a symmetry   
of the IH geometry $(q,\D)$. If  $(\IH, [\ell])$ is a vacuum IH, 
then \cite{abl-g} there is another vector field $\ell'$ 
tangent to $\IH$, such that $(\IH, [\ell'])$ is an extremal 
null-symmetric IH characterized above. We generalize now this result  
to the electrovac case. (Our proof will also be valid for the vacuum case. 
In this way we fill a small gap in the proof sketched in 
\cite{abl-g}.) The vector field $X$ can be written as 
\begin{equation}  
  X\ =\ f\ell 
\end{equation} 
where the function $f$ satisfies the equations \eqref{nulsym}. 
Those equations imply that 
\begin{equation}\label{eq:f} 
  f \ =\ b e^{-\kl v} + a_0 
\end{equation}  
where $a_0$ is a constant, $\kl\not=0$ is the surface gravity of $\ell$,  
\begin{subequations}\label{eq:_nexnull1}\begin{align} 
  \opD v\ &=\ 1,  & 
  \opD b\ &=\ 0,  \tag{\ref{eq:_nexnull1}}  
\end{align}\end{subequations} 
and the function $b$ satisfies the following equations 
 \begin{subequations}\label{eq:nonuniq}\begin{align} 
  \label{eq:nu1} 
    \left[ \frac{1}{2}\left(\m\mb+\mb\m - (\alpha-\bar{\beta})\m 
                            - (\bar{\alpha}-\beta)\mb \right) 
          +\pi\m+\bar{\pi}\mb+\mu\kl \right] b &= 0, \\ 
  \label{eq:nu2} 
    \left[ \left( \m+\bar{\alpha}-\beta+2\bar{\pi} \right)\m  
          +\bar{\lambda}\kl \right] b &= 0.   
\end{align}\end{subequations} 
 
The resulting vector field satisfies  
\begin{equation} 
  \D_{X}X\ =\ a_0 \kl X. 
\end{equation} 
Therefore, if $b$ nowhere vanishes, then $X$ corresponding
to $a_0=0$ defines an extremal IH on $\IH$. We show the non-vanishing 
of the function $b$ assuming that $(\IH, [\ell])$ admits an electromagnetic 
field $F$  of the properties which make  $\IH$ an electrovac
IH, and such that  
\begin{equation}\label{XF}  
  \Lie_{f\ell}(F_{\mu}{}^{\nu} e^{\mu} \otimes e_{\nu}) \ =\  0.  
\end{equation}   
  
The integrability condition for the equations \eqref{eq:nonuniq} 
has the following form 
\begin{equation}\label{eq:nonuniq_int} 
  (3\Psi_2-R_{m\bar{m}})\mb b + \left((\mb +3\pi)\Psi_2\right)b = 0.  
\end{equation} 
whereas \eqref{XF} implies 
\begin{equation}\label{eq:Dphi2}  
  \Phi_2 \opD f - 2\Phi_1 \mb f\ =\ 0.  
\end{equation} 
The part of Maxwell equations used in definition of electrovac isolated horizon 
allows to express $\Phi_2$ by $\Phi_1$: 
\begin{equation} 
  \kl\Phi_2 \ = \ \mb\Phi_1 + 2\pi\Phi_1.  
\end{equation} 
Then, equation \eqref{eq:Dphi2} for $f$ given by \eqref{eq:f} becomes 
\begin{equation}\label{eq:Fconstr} 
  (\mb\Phi_1 + 2\pi\Phi_1)b + 2\Phi_1\mb b \ = \ 0. 
\end{equation} 
This constraint can easily be integrated and is equivalent to 
\begin{equation}  
  \Phi_1 b^2 e^{-2iU} \ = \ E_0 = \const  
\end{equation} 
 
If $E_0$ is not zero then $b$ cannot vanish, therefore we assume that  
\begin{equation} 
  E_0\ =\ 0.  
\end{equation}    
Let  
\begin{equation} 
  {\cal U}\ =\ \{x\in \IH: b(x)\not= 0 \}. 
\end{equation}  
Since  
\begin{equation} 
  {\Phi_1}|_{\cal U}\ =\ 0  
\end{equation}   
the condition \eqref{eq:nonuniq_int} implies
that the following complex-valued function $C$ given by 
\begin{equation}  
  C\ :=\ \Psi_2 b^3 e^{-3iU}, 
\end{equation} 
is constant on $\cal{U}$.  
Suppose $C\neq 0$ on a connected component of $\cal{U}$.
If it is not the entire sphere $\hIH$, then  
there is a sequence $x_n\in{\cal U}$, $n=1,2,3,...$ 
such that  
\begin{equation} 
  x_n \rightarrow x_0 \quad \text{and} \quad b(x_0) = 0. 
\end{equation} 
But then  
\begin{equation} 
  |\Psi_2(x_n)| \ \rightarrow \ \infty.  
\end{equation} 
Hence, the only possibility for $b$ to vanish somewhere on $\IH$ is    
\begin{subequations}\label{eq:_nexnull2}\begin{align} 
  C|_{\cal{U}} \ &= \ 0,  & 
  {\Psi_2}|_{\cal U} \ &= \ 0. \tag{\ref{eq:_nexnull2}} 
\end{align}\end{subequations} 

In the non-extremal case we can always choose a foliation by 
space-like cross sections, preserved by $\ell$, such that 
\begin{equation}
\pi =\ -i\bar{\delta}U.
\end{equation}
Given that foliation, consider on $\hIH$ a 1-form  
\begin{equation} 
  \beta \ := \ \partial \left( e^{iU} b \right)  
  \ := \ \frac{\partial}{\partial z} \left( e^{-2iU} b \right) dz 
\end{equation} 
where $z$ is any local holomorphic coordinate on $\hat{\IH}$.  
Then the vanishing of $\Psi_2$ and  equation \eqref{eq:nu1} 
imply 
\begin{equation} 
  \bar{\partial}(e^{-2iU}\beta)\ =\ 0 
\end{equation} 
{\it everywhere} on $\hat{\IH}$. On a sphere, the only 
solution to  this equation is  
\begin{equation}  
  \beta \ = \ 0 \ = \ \partial \left( e^{iU} b \right). 
\end{equation} 
And that implies 
\begin{equation} 
  b\ =\ \const 
\end{equation} 
on the entire $\IH$. Since we assumed $[(b e^{-\kl v}+a_0)\ell]\neq [\ell]$, 
the conclusion is that 
\begin{equation} 
  b(x)\ \neq\ 0 \quad \text{for every} \quad x \ \in \ \IH,  
\end{equation} 
and the vector field  
\begin{equation} 
  \ell'\ =\ b e^{-iv} \ell 
\end{equation} 
defines an extremal IH on $\IH$.  
Our discussion is summarized by the following: 

\begin{prp}\label{thm:ExNull} 
Suppose a non-extremal IH $(\IH, [\ell])$ admits 
a null vector field $X\notin [\ell]$ which generates
a symmetry of the IH geometry. If $\IH$ admits
an electromagnetic field $F$ which makes it an electrovac
IH, and such that  
\begin{equation} 
        \Lie_{X}(F_{\mu}{}^{\nu} e^{\mu} \otimes e_{\nu})_{(\IH)} \ = \ 0, 
\end{equation} 
then $(\IH, [X])$ is  an extremal IH. 
\end{prp} 
       
\section{General axisymmetric solution: the uniqueness of the 
monopolar extremal Kerr-Newman IH}\label{sec:axsol} 
In this section, we consider an axi-symmetric extremal IH
$(\IH, [\ell])$ and an electromagnetic field $F$ such that
\begin{equation}
  \Lie_{\bf{\Phi}} \left(F-i\star F\right)_\nu{}^\mu e^\nu_{(\IH)}
\otimes e_\mu \ =\ 0, 
\end{equation} 
where $\bf{\Phi}$ is a generator of the axial symmetry of $\IH$. 
We solve completely the constraint equations \eqref{GT}-\eqref{LielF}
imposed on the geometry $(q,\D)$  and on the 
pullback $F_{(\IH)}$.
First, we  find all the local solutions. Next we select those that can  
be defined globally on $\IH = \hIH\times\re$. Finally, we identify the 
solutions 
by using their area, and electric and magnetic charges
as corresponding to those defined by the extremal Kerr-Newman solution
(of the electromagnetic field modified by the allowed magnetic charge).    

\subsection{General solution of the constraints} 
\noindent{\bf The local constraints.} 
 In the  coordinates $(\varphi, x)$, the null 2-frame on $\hIH$ takes the 
following form: 
\begin{align}\label{P} 
  \hat{m}\ &=\ \frac{1}{2}\left(\frac{1}{P}\partial_x 
                                +iP\partial_\varphi\right), & 
  \hat{e}^1\ &=\ \left(Pdx - i \frac{1}{P}d\varphi\right). 
\end{align} 
Due to the rotational symmetry, the function $P$ and all the 
functions used in the constraint equations are functions of the 
variable $x$ only. For this frame, the connection coefficient $a$ 
of \eqref{2gamma} is 
\begin{equation}
  a\ =\ \frac{P_x}{4P^2}. 
\end{equation} 
 
The constraints \eqref{112} on the functions 
$P$, $U$, $B$, $\Phi_1$ read 
\begin{subequations}\label{1b}\begin{align} 
  \label{12b} 
    \left(\frac{B}{P}\right)_{xx} + (P_x)^2\frac{B}{P} 
    + 8 P\frac{|E_0|^2}{B^3} \ &=\ 0 \\ 
  \label{11b} 
    B_{xx} - (U_x)^2 B\ &=\ 0, &  
    (U_xB^2)_{x}\ &= 0,         \tag{\ref{1b}bc} 
\end{align}\end{subequations} 
where the second and the third equations above, are the real and the imaginary 
parts of the constraint \eqref{11} (the third one being extra  multiplied by 
$B$). The gauge freedom is 
\begin{subequations}\label{eq:_int1}\begin{align}
  U \ &\mapsto \ U + U_0,  & 
  B \ \mapsto  \  B_0 B,  \tag{\ref{eq:_int1}} 
\end{align}\end{subequations}
whereas the freedom in the choice of the coordinates is 
already eliminated by the choice of the coordinate $x$. 
\medskip 
 
\noindent{\bf The global existence conditions.}  
The global existence conditions are that the  metric tensor 
\begin{equation}
  \hq \ := \ \hat{e}^1\otimes \hat{e}^2 +  \hat{e}^2\otimes \hat{e}^1 
\end{equation} 
derived from the  function $P$ and the scalar functions $U$, $B$, 
$\Phi_1$ are globally defined on $\hIH$ and differentiable as many 
times as one needs, and that $B$ are nowhere zero on $\hIH$ and the metric 
tensor $\hq$ is nowhere degenerate. 
 
In the  consideration below, the global existence conditions will be applied in the 
following way. We will first find local  solutions of the constraint equations
\eqref{112}; subsequently
we will find $B$, $U$, $\Phi_1$ and $P$ well defined 
everywhere on $\hIH$ except the poles $x=x_\pm:= \pm\frac{A}{8\pi}$. 
Next, we will apply the following  condition for the coordinate 
function $P$ at the poles, necessary for the metric $\hq$ to be 
of the class $C^1$, 
\begin{subequations}\label{glob}\begin{align} 
  \lim_{x\to x_\pm}\frac{1}{P}\ &=\ 0,\\ 
  \lim_{x\to x_\pm}\left(\frac{1}{P^2}\right)_x\ &=\ \mp 2. 
\end{align}\end{subequations} 
\medskip 
 
\noindent{\bf Integration.}  
We start with the equations 
\eqref{11b}. The integration of the second one gives 
\begin{equation}\label{tOmega} 
  U_x B^2\ =\ \tOmega\ = \const. 
\end{equation} 
Note that $\tOmega=0$ if and only if the horizon $\IH$ is 
non-rotating, 
\begin{equation}
  \tOmega = 0\ \Leftrightarrow\ d\hat{\omega}=0\  \Leftrightarrow\ 
  \ d\omega=0. 
\end{equation} 
Consider the rotating case first, 
\begin{equation}
  \omega\neq 0. 
\end{equation} 
We use \eqref{tOmega} to eliminate the function $U$ from the first 
equation in \eqref{11b}, which now reads 
\begin{equation}
  B_{xx} - \frac{\tOmega^2}{B^3}\ =\ 0. 
\end{equation} 
The general solution for $B$ is 
\begin{equation}
  B\ =\ B_0\left(\Omega^2 + (x-x_0)^2\right)^{\frac{1}{2}},\qquad \Omega\ =\ 
  \frac{\tOmega}{B_0{}^2}, 
\end{equation} 
where $B_0$ and $x_0$ are arbitrary constants. Using the integral 
\eqref{tOmega} again we find $U$, 
\begin{equation}
  U\ =\ \arctan\left(\frac{x-x_0}{\Omega}\right) + U_0,\ \ U_0=\const. 
\end{equation} 
 
In this way we have exhausted the vanishing of the Ricci tensor component 
$R_{mm}$, and found  both of the potentials defining the rotation 
1-form potential $\omega$. But to know the 1-form $\omega$ itself we 
need the frame operator given by $P$. To find $P$ we turn to 
the equation \eqref{12} with the above solutions substituted 
for $U$ and  $B$ and $\Phi_1$ given by \eqref{phi}, 
\begin{equation}
  \begin{split} 
    \left[\frac{d^2}{dx^2} + \frac{2(x-x_0)}{(x-x_0)^2 + \Omega^2}\frac{d}{dx} 
    + \frac{4\Omega^2}{((x-x_0)^2+\Omega^2)^2} \right]\frac{1}{P^2}\ =\\ 
    - \frac{16|E_0|^2}{B_0{}^4((x-x_0)^2+\Omega^2)^2}. 
  \end{split} 
\end{equation} 
 
The general solution is 
\begin{equation}
  \frac{1}{P^2}\ =\ \frac{c_1(\Omega^2 - (x-x_0)^2) + 2c_2\Omega(x-x_0) 
  -\frac{8|E_0|^2}{B_0{}^4}}{\Omega^2 + (x-x_0)^2}. 
\end{equation} 
 
\noindent The requirement, that $P$ be finite on the open interval 
$x\in\, ]-\frac{A}{8\pi}, \frac{A}{8\pi}[$, and satisfy the 
conditions \eqref{glob} at the poles determines the 
values of the constants $x_0, c_1, c_2, \frac{|E_0|^2}{B_0{}^4}$. 
\medskip 
 
\noindent{\bf The solution.} In the rotating case, the resulting 
solutions $P, U, B, \Phi_1$ of the Maxwell-Einstein equations 
\eqref{mbphi}, \eqref{112} on $\IH$ can be expressed by three real 
parameters $A, \alpha$ and $\theta_0$ of the domain of dependence 
specified below, where the third one is a constant phase of 
$\Phi_1$ and  {\it does not affect the horizon geometry 
$(q,\D)$}: 
\begin{subequations}\label{data_rot}\begin{align} 
  \label{P2} 
    P^2(x)\ &=\ \frac{4\pi(1+\alpha^2)}{A}\ 
                \frac{A^2+(8\pi)^2\frac{1-\alpha^2}{1+\alpha^2}x^2} 
                     {A^2-(8\pi)^2x^2},\\ 
  U\ &=\ \pm\arctan\left(8\pi\sqrt{\frac{1-\alpha^2}{1+\alpha^2}}\, 
         \frac{x}{A} \right),\\ 
  B\ &=\ \left(1 + \frac{1-\alpha^2}{1+\alpha^2} \frac{(8\pi x)^2}{A^2}  
           \right)^{\frac{1}{2}},\\ 
  \Phi_1\ &=\ e^{i\theta_0}  
     \frac{2\sqrt{\pi}A^{\frac{3}{2}}\alpha}{1+\alpha^2}\ 
     \frac{\left(A^2-\frac{1-\alpha^2}{1+\alpha^2}(8\pi x)^2\right) 
           \pm 2iA\sqrt{\frac{1-\alpha^2}{1+\alpha^2}}(8\pi x)} 
          {\left(A^2+\frac{1-\alpha^2}{1+\alpha^2}(8\pi x)^2\right)^2}, 
\end{align}\end{subequations} 
\begin{subequations}\label{xxx_par}\begin{align} 
  \alpha &\in [0, 1[, & 
  A &\in ]0,\infty[, & 
  \theta_0 &\in [0,2\pi[. \tag{\ref{xxx_par}} 
\end{align}\end{subequations} 
Above we have fixed the gauge freedom in the rescaling of $B$ by a 
constant and adding a constant to $U$ such that as $\alpha\rightarrow 1$ 
the function $B$ has a finite limit and $U$ goes to zero. 
 
In the non-rotating case, we take the constant $\tOmega$ 
in \eqref{tOmega} to be zero, 
\begin{equation}
  \tOmega\ =\ 0, 
\end{equation} 
and we repeat all the steps of the derivation above including 
solving the conditions \eqref{glob} at the poles. The result is a 
family of solutions parametrized by a real constant $A>0$, 
\begin{subequations}\label{data_nrot}\begin{align} 
  P^2(x)\ &=\ \frac{8\pi A}{A^2-(8\pi x)^2}\\ 
  U\ &=\ 0 \\ 
  B\ &=\ 1 \\ 
  \Phi_1\ &=\ e^{i\theta_0}\left(\frac{\pi}{A}\right)^{\frac{1}{2}}. 
\end{align}\end{subequations} 
It turns out that the non-rotating solutions coincide with $P$, 
$U$, $B$, $\Phi_1$ obtained by substituting the parameter $\alpha$ 
in the rotating case for the excluded values $\pm 1$. (Locally, the 
non-rotating case has solutions then those obtained by the limit 
$\alpha\rightarrow 1$.) 
 
\subsection{The uniqueness of the extremal Kerr-Newman IH}
\label{sec:results} 
The solutions $(q,\omega, 
F_{(\IH)})$ are given by \eqref{G43}, \eqref{homega}, \eqref{pi}, \eqref{F}, 
\eqref{Phi0}, \eqref{P}, \eqref{data_rot}, \eqref{data_nrot} and set up 
a three-parameter family. To compare the 
solutions with those given by the Kerr-Newman metrics, we use the 
horizon area, and the electro magnetic charges. The electric and 
magnetic charges are given by the real and, respectively, the 
imaginary part of the following integral: 
\begin{equation}\label{charge} 
  \frac{1}{4\pi}\int_{\hIH} *F + i F \ =\ 
    e^{i\theta_0}\alpha\sqrt{\frac{A}{4\pi}}. 
\end{equation} 
The condition that magnetic charge vanishes implies 
\begin{equation}
  e^{i\theta_0} \ =\ \pm 1.  
\end{equation} 
Therefore, in the absence of the magnetic charge, 
the derived family of $(q,\omega, F_{(\IH)})$ is globally 
and  uniquely parametrized by the area $A$ of  (a slice of) the horizon 
and the electric charge $Q$ contained therein, having the following domain 
of dependence, 
\begin{subequations}\label{AQ}\begin{align} 
  A \ &\in \ ]0, \infty], & 
  Q \ &\in \ [-\sqrt{\frac{A}{4\pi}}, \sqrt{\frac{A}{4\pi}}]. 
    \tag{\ref{AQ}} 
\end{align}\end{subequations} 
But for each pair $(A,Q)$ of \eqref{AQ} there {\it is} the 
extremal Kerr-Newman metric and the corresponding electromagnetic 
field defining an isolated horizon of the area $A$ and the 
electric charge $Q$. Therefore, our family of $(q, \omega, 
F_{(\IH)})$ consist of the ones given by the extremal Kerr-Newman 
solutions.. 
In conclusion:
\medskip 
 
\begin{thm}\label{uniq} Suppose  $(\IH, [\ell])$ is an extremal 
isolated horizon and $F_\mu{}^\nu e^\mu_{(\IH)}\otimes e_\nu$ is 
an electromagnetic field defined on $\IH$, such that 
the flow $[\ell]$ is a symmetry of the electromagnetic field
and the total magnetic charge of $\IH$ is zero.
Suppose also, that both the IH geometry and the electromagnetic 
field admit a symmetry group of motions of $\IH$ isomorphic to O(2).
Then, the Einstein-Maxwell constraints 
\eqref{GT}-\eqref{mF}
are satisfied if and only if, the induced metric 
$q$, the rotation potential 1-form  $\omega$ and the pullback of 
the electromagnetic field 2-form  $F_{(\IH)}$ coincide
with those defined by the extremal Kerr-Newman solution on the 
event horizon. Admitting a non-zero magnetic charge does not
affect the possible geometries $(q,\D)$. 
\end{thm} 

Recall, that after solving the constraints,
the remaining components of the induced connection $\D$, that is 
the transversal expansion $\mu$, and transversal shear $\lambda 
(e^2\otimes e^2)_{(\IH)}$ of the lives of a preserved by the 
symmetries  foliation, are arbitrary real function and complex 
valued tensor, respectively, satisfying the symmetry conditions 
\begin{subequations}\begin{align} 
  \ell^a\partial_a \mu\ =\ &0\ =\ \bf{\Phi}^a\partial_a \mu\ \\ 
  \Lie_\ell (\lambda e^2\otimes e^2)_{(\IH)}\ =\ &0\  
     =\ \Lie_{\bf{\Phi}}(\lambda e^2\otimes e^2)_{(\IH)}. 
\end{align}\end{subequations} 
Similarly, the component $\Phi_2$ of the electromagnetic 
field has to satisfy the symmetry assumptions, whereas the 
Einstein-Maxwell equations do not impose any extra constraint on 
it at $\IH$. Despite of this arbitrariness, the symmetries of 
those remaining components affected the form of the constraints 
which were solved above.   
For the completeness, the full extremal Kerr-Newman geometry 
$(q,\D)$ is given by the following $\mu$ and $\lambda$, 
\begin{subequations}\begin{align} 
  \mu \ &=\ -\ \frac{ \sqrt{\frac{8\pi}{1+\alpha^2}} A^{\frac{3}{2}} } 
                    { A^2 + \frac{1-\alpha^2}{1+\alpha^2} (8\pi x)^2 } \\ 
  \lambda \ &=\ -\ \frac{ \sqrt{\frac{8\pi}{1+\alpha^2}} A^{\frac{1}{2}} 
                   \frac{1-\alpha^2}{1+\alpha^2} \left(A^2-(8\pi x)^2\right) } 
              { \left( A^2 + \frac{1-\alpha^2}{1+\alpha^2}(8\pi x)^2 \right) 
                \left( A  
                   \mp i\sqrt{\frac{1-\alpha^2}{1+\alpha^2}}(8\pi x) \right) 
              }  \ , 
\end{align}\end{subequations} 
modulo  the choice of the foliation of $\IH$.

\section*{Acknowledgments} 
 
We would like to thank Abhay Ashtekar, Jiri Bicak, Piotr Chrusciel, 
Jim Isenberg, Jacek Jezierski, Vince Moncrief, and Bernd Schmidt for 
discussions and Josh Willis for careful reading the paper. 
 This work was 
supported in part by the Polish Committee for Scientific Research 
(KBN) under grant no 2 P03B 12724, the Eberly research funds of Penn State 
and the Albert Einstein Institute of the Max Planck Society.

\appendix 
 
\section*{Appendix. The intrinsic Hodge dual} 
 
The intrinsic Hodge dual we used in \eqref{mF} is defined on $\IH$  
as follows.  
An orientation and time orientation of spacetime 
are used to define an orientation of space-like 2-surfaces 
of $\IH$. A neighbourhood of a 2-surface $\hIH$ is diffeomorphic to  
$\hIH\times\re\times\re$ where the factors $\re$  correspond to 
the  null geodesics  orthogonal to $\hIH$: the first factor  the ones 
transvers to $\IH$ and the second to those contained in $\IH$. The factors 
$\re$ are oriented according to the time orientation. 
Then the orientation of $\hIH$ is such that the orientation of 
$\hIH\times\re\times\re$ agrees with the orientation of spacetime. 
Next, we introduce on $\IH$ the 2-volume form \cite{abf} ${}^{(2)}\epsilon$, 
a 2-form such that the area of any 2-surface $\hIH\subset \IH$ 
is $\int_{\hIH} {}^{(2)}\epsilon$. 
Given a 1-form $\mu$ defined on $\IH$, such that $\ell^a\mu_a=0$,  
we define  
\begin{equation} 
  {}^{(2)}*\mu_a\ =\ {}^{(2)}\epsilon_{ab}q^{bc}\mu_{c}, 
\end{equation}  
where $q^{ab}$ is such that $q_{ab}q^{bc}\nu_c=\nu_a$ for every  
1-form $\nu$ orthogonal to $\ell$.

\end{document}